# Frequency ratio of the $^{229m}$Th nuclear isomeric transition and the $^{87}$Sr atomic clock

**Chuankun Zhang[1,*], Tian Ooi[1], Jacob S. Higgins[1], Jack F. Doyle[1], Lars von der Wense[1], Kjeld Beeks[2], Adrian Leitner[2], Georgy A. Kazakov[2], Peng Li[3], Peter G. Thirolf[4], Thorsten Schumm[2], and Jun Ye[1,*]**

**[1]JILA, NIST and University of Colorado, Department of Physics, University of Colorado, Boulder, CO 80309**
**[2]Vienna Center for Quantum Science and Technology, Atominstitut, TU Wien, 1020 Vienna, Austria**
**[3]IMRA America, Inc., 1044 Woodridge Ave., Ann Arbor, MI 48105, USA**
**[4]Ludwig-Maximilians-Universität München, Garching, Germany**

**[*]Corresponding author email address: chuankun.zhang@colorado.edu, ye@jila.colorado.edu**

**Optical atomic clocks[1,2] use electronic energy levels to precisely keep track of time. A clock based on nuclear energy levels promises a next-generation platform for precision metrology and fundamental physics studies. Thorium-229 nuclei exhibit a uniquely low energy nuclear transition within reach of state-of-the-art vacuum ultraviolet (VUV) laser light sources and have therefore been proposed for construction of the first nuclear clock[3,4]. However, quantum state-resolved spectroscopy of the $^{229m}$Th isomer to determine the underlying nuclear structure and establish a direct frequency connection with existing atomic clocks has yet to be performed. Here, we use a VUV frequency comb to directly excite the narrow $^{229}$Th nuclear clock transition in a solid-state $CaF_2$ host material and determine the absolute transition frequency. We stabilize the fundamental frequency comb to the JILA $^{87}$Sr clock[2] and coherently upconvert the fundamental to its 7$^{th}$ harmonic in the VUV range using a femtosecond enhancement cavity. This VUV comb establishes a frequency link between nuclear and electronic energy levels and allows us to directly measure the frequency ratio of the $^{229}$Th nuclear clock transition and the $^{87}$Sr atomic clock. We also precisely measure the nuclear quadrupole splittings and extract intrinsic properties of the isomer. These results mark the start of nuclear-based solid-state optical clock and demonstrate the first comparison of nuclear and atomic clocks for fundamental physics studies. This work represents a confluence of precision metrology, ultrafast strong field physics, nuclear physics, and fundamental physics.**

Time and frequency are the most precisely measured physical quantities in experimental physics[1,5]. Employing ultranarrow linewidth electronic transitions in the visible spectral domain, current optical atomic clocks have achieved measurement precision better than $1 \times 10^{-20}$ and systematic uncertainty at $8 \times 10^{-19}$ in fractional frequency units[2]. These systems are poised to explore fundamental physics such as measurements of quantum effects under gravity and searches for new physics beyond the standard model[6].

The tremendous progress of optical atomic clocks traces its scientific roots to increasingly higher quality factors of natural resonances, highly coherent lasers, precise quantum state control of relevant atomic degrees of freedom, understanding of complex many-body physics, and strong connections to fundamental physics. These important ingredients form the foundation of the next generation of clocks. Harnessing natural resonances in the VUV spectrum and beyond while maintaining long coherence times naturally leads to higher quality factors. A clock based on a fundamentally new platform could greatly advance tabletop experiments' ability to explore new physics. Another increasingly important need is to take clocks out of the laboratory without adversely impacting their precision or accuracy.

It is with this scientific context that a low-lying thorium-229 nuclear transition has emerged as a highly desirable candidate for the next generation clock[3,4]. Remarkably, the metastable isomeric state $^{229m}Th$ is only 8.4 eV (~148 nm) higher in energy than the ground state, with a lifetime[7–9] of $\sim 10^3$ s. This nuclear transition is lower in frequency than almost every other known nuclear transition by several orders of magnitude due to a nearly exact cancelation of the MeV level nuclear energy terms. Therefore, precise laser spectroscopy of the $^{229}Th$ nuclear transition dramatically enhances the sensitivity of tests of fundamental physics[8,10,11], such as searches for ultralight dark matter candidates, or temporal variations of the fine structure constant and the dimensionless strong interaction parameter. It has also been proposed for study of fundamental electron-nuclear coupling interactions such as the electron bridge process[12]. Furthermore, the nuclear transition is highly insensitive to external electromagnetic perturbations[3], making it an ideal portable clock.

Over the past two decades the determination of the energy and properties of $^{229m}Th$ has steadily advanced, initially indirectly[13–16] and most recently from a flurry of nuclear physics experiments[17–21] after the first detection of the internal conversion decay[22] of $^{229m}Th$. The first direct observation of the radiative decay[23] of $^{229m}Th$ was reported from CERN only a year ago, using $^{229}Ac$ implanted into a $CaF_2$ crystal. Upon beta decay, a large fraction of the daughter isotope branched into the $^{229m}Th$ state, and their subsequent radiative decay photons were detected on a VUV spectrometer. This measurement placed the transition frequency uncertainty at the THz level and reported a radiative decay half-life of 670(102) s, corresponding to a lifetime of 967(147) s. Here and throughout the numbers in parenthesis represent the 1-σ uncertainty. Later, a group from Okayama University used X-rays to pump a Th-doped $CaF_2$ crystal to the second excited state[24] of 29 keV. which partially decayed to $^{229m}Th$ and yielded a radiative decay with a half-life of 447(25) s (lifetime 645(36) s). Another recent experiment at RIKEN successfully trapped isomeric $^{229m}Th^{3+}$ (populated via $^{233}U$ decay) in an ion trap[25] and measured a half-life of $1400^{+600}_{-300}$ s (lifetime $2020^{+866}_{-433}$ s).

The first resonant excitation was reported at PTB using a $^{229}Th$-doped $CaF_2$ crystal irradiated by a nanosecond pulsed VUV laser generated via four-wave mixing[26]. This result was soon independently verified in a $^{229}Th$-doped $LiSrAlF_6$ crystal at UCLA using a similar four-wave mixing light source[27]. These two measurements observed a single resonance profile with a laser-limited linewidth on the order of 10 GHz and placed the uncertainty of the transition at a similar frequency scale. The fluorescent lifetimes reported in the $CaF_2$ and $LiSrAlF_6$ crystals were 630(15) s and $568(13)_{stat}(20)_{sys}$ s respectively. The different isomer lifetimes observed in these experiments can be attributed to the Purcell effect, where a higher refractive index of the crystalline environment leads to a higher density of photon states[28].

In this work, we follow the early proposals[29,30] and embed $^{229}$Th in a solid-state $CaF_2$ crystal. We probe the narrow nuclear clock transition directly with a single line of a VUV frequency comb. The VUV comb is generated via a coherent high harmonic generation process and is frequency stabilized to the JILA $^{87}$Sr optical clock[2]. In the demonstrated precision spectroscopy of the $^{229}$Th nuclear clock transition, we directly resolve the narrow line structure arising from the interaction between the nuclear electric quadrupole moments and the crystal electric field gradient. We determine the absolute transition frequency to the kHz level. This work establishes a milestone for direct frequency connection between a nuclear clock and an optical atomic clock, provides an enhanced determination of important parameters of the nucleus and the crystal field, and paves the way to search for temporal variations of fundamental constants and to develop a portable nuclear clock based on $^{229}$Th-doped crystals.

## Direct frequency link between the $^{87}$Sr atomic clock and the $^{229m}$Th isomeric transition

The JILA Sr lattice clock[2] serves as the absolute frequency reference for our fundamental comb in the infrared (IR), see Fig. 1a. A local oscillator at 698 nm, which serves as the Sr clock laser, is stabilized to a cryogenic silicon cavity[31] operated at 1.5 μm via a separate optical frequency comb, inheriting excellent frequency stability. The absolute frequency of the Sr clock laser is periodically calibrated with the JILA Sr lattice clock to correct for long-term drifts. While we did not actively perform a systematic evaluation of the Sr clock, its overall uncertainty is estimated to be below $1 \times 10^{-16}$ across the period of our measurements[32], providing a sufficiently accurate frequency reference for the $^{229}$Th transition frequency determination.

Frequency combs are ideal tools to transfer optical frequency references across large wavelength ranges[33]. Their strictly equidistant comb lines permit precise and direct readout of optical frequencies, while their femtosecond pulsed nature in the time domain allows efficient nonlinear frequency conversion[34] to different spectral domains. To connect the absolute frequency of the $^{87}$Sr clock laser to the $^{229}$Th nuclear transition frequency, we employ a VUV comb generation process.

Frequency combs in the vacuum-to-extreme ultraviolet have been under continuous development over the past two decades[35–39]. Our specific VUV frequency comb apparatus is described in a previous publication[40]. We start with a Yb fiber-based oscillator[41] followed by nonlinear preamplification and chirped pulse power amplification with a large mode area Yb-doped fiber to reach an average output power of 40-50 W. We use a femtosecond enhancement cavity for coherent enhancement of the IR comb, obtaining an intracavity train of femtosecond pulses with an average power of ~5-7 kW and pulse durations of ~150-200 fs. The intracavity IR pulse is focused (peak intensity ~$10^{13}$ W/cm$^2$) into a xenon gas jet for efficient 7$^{th}$ harmonic generation. To ensure our 7$^{th}$ harmonic centers around 148.3 nm, we bandpass filter the seed spectrum before the final Yb-fiber amplifier to a center wavelength of 1038 nm. The VUV comb spectral envelope width is ~1 nm, fully covering the known uncertainty range of the nuclear transition. A high repetition rate of 75 MHz is chosen to preserve the optical phase coherence and generate well-separated comb lines for precision spectroscopy.

To prepare for high resolution spectroscopy and clock operation on the nuclear transition, we have made several critical improvements to our VUV frequency comb apparatus. We employ a grazing incidence plate[42,43] to outcouple the VUV comb with ~50% efficiency. We translate the plate to a new spot every ~10 hours to avoid VUV-induced plate degradation. A home-built xenon recycling system is installed to reduce the operational cost from using a continuous gas jet. By engineering for excellent passive thermal stability, we achieve stable, continuous, full-power operation of the VUV comb for many days, during which only occasional relocking and maintenance of the laser is needed.

The outcoupled 7$^{th}$ harmonic is selectively reflected to the $^{229}$Th target using a multilayer coated mirror. Additionally, we take advantage of the chromatic dispersion in a $MgF_2$ lens to further filter out other harmonics geometrically. We estimate the available power in the 7$^{th}$ harmonic on the target to be ~200 μW, corresponding to 1 nW per comb mode.

**Spectroscopy setup**

A $^{229}$Th:$CaF_2$ single crystal with a doping concentration of $5 \times 10^{18}/cm^3$ was grown at TU Wien using a miniaturized vertical gradient freeze method followed by a fluorination treatment[44,45]. A small piece ($1.8 \times 0.7 \times 1.4$ mm) of this $^{229}$Th:$CaF_2$ crystal is cut and polished for this experiment. Fluorescence from the $^{229m}$Th isomer has been observed in crystals cut from the same ingot at both PTB[26] and Okayama University[24].

The crystal is mounted at the focus of a parabolic collection mirror and cooled down to 150(1) K (see Methods). We illuminate the crystal with our VUV beam (0.2 mm diameter) along the 1.4 mm length direction. Fluorescence photons collimated by the parabolic mirror are selectively steered with a series of multilayer coated filters ($150 \pm 10$ nm) to a photomultiplier tube (PMT) for single photon counting. Fig. 1b shows a conceptual visualization and a photo of the setup. The $CaF_2$ crystal scintillates, tracing out our VUV beam path as the thin line shown in the photo. The transmitted laser is terminated at the yellow Ce:YAG fluorescent plate. The effective efficiency of the setup, from the nucleus emitting a photon to a count collected on the PMT, is estimated to be 0.3%.

**Comb locking and full range scan**

For comb stabilization, a supercontinuum of the IR comb is generated in a highly nonlinear fiber with a picked off portion of the pre-amplified light. The supercontinuum light is used in the *f–2f* referencing scheme to stabilize the carrier-envelope offset frequency $f_{CEO}$ to a radiofrequency reference. The $f_{CEO}$ lock point is chosen to be –8 MHz to match the dispersion in the femtosecond enhancement cavity[35,36]. We phase lock one of the supercontinuum comb lines to the $^{87}$Sr clock laser with an offset frequency $f_{beat}$ set by a direct digital synthesizer (DDS). Changing $f_{beat}$ while fixing $f_{CEO}$ allows us to precisely tune the comb repetition rate $f_{rep}$. Additionally, a narrow linewidth Mephisto laser at 1064 nm is used as a short-term reference to stabilize the comb linewidth, whose frequency is slowly steered to maintain the phase lock between the supercontinuum comb and the Sr clock laser. The Mephisto laser also serves as an auxiliary laser for stabilizing the enhancement cavity length (see Methods).

The $7^{th}$ harmonic has a frequency comb structure with the same repetition rate as the fundamental, but an offset frequency of $7 \times f_{\mathrm{CEO}}$ and integer comb mode numbers $N \approx 2.7 \times 10^7$. To search for the transition, we scan $f_{\mathrm{rep}}$ in the range of approximately 2.8 Hz to scan the comb structure underneath the $7^{th}$ harmonic spectral envelope without changing the envelope itself. This scan range is multiplied by the large comb mode number $N$ in the VUV domain, making the $N^{th}$ comb mode's frequency at the end of scan overlap the starting point for the $(N+1)$ comb mode frequency (Fig. 2, top panel). Thus, we fully cover the spectrum under the ~1 nm wide comb spectral envelope.

A typical experimental cycle consists of two parts: irradiation and photon detection. We start by irradiating the crystal for 400 s, during which the PMT is turned off to avoid damage from scattered VUV photons. The comb frequencies are swept with a sawtooth profile over a 240 kHz scan range in the VUV. The starting frequency of the sawtooth is stepped by 240 kHz between cycles. During photon detection, the IR beam is diverted from the cavity by an AOM. This completely turns off the VUV comb, removing any background from scattered light. We turn on the PMT to count individual fluorescent photons from the $^{229m}$Th isomer decay for 200 s. The photon counts in each 1 s time bin is collected using a digital counter.

Fig. 2 shows the results of a full range scan. In the middle panel, we plot the photon count rate averaged over the 200 s detection window as a function of $f_{\mathrm{rep}}$. The background photon count rate of ~415 counts per second comes from the intrinsic radioactivity of the $^{229}$Th:$CaF_2$ crystal[46]. On top of the background, six clear spectroscopic features are observed (highlighted in blue/green). In the bottom panel, we plot the count rate as a function of time after binning $f_{\mathrm{rep}}$ to ~15 mHz bins (~400 kHz bin size in VUV frequency) to observe the time-dependent fluorescence for each scan step. We clearly see that the fluorescence from each feature persists beyond the 200 s detection window, consistent with the long lifetime[26] of $^{229m}$Th embedded in $CaF_2$. We take the four strongest peaks (highlighted in green) from the scan to perform absolute frequency determination measurements. The two weak peaks may originate from different electronic environments in the crystal and will be subject to future studies.

**Lineshape and center frequency determination**

To measure the lifetime of the $^{229m}$Th state, we place one comb line on resonance and irradiate the crystal for 1200 s. We then monitor the fluorescence count rate for ~90 minutes. The trace in Fig. 3a follows a clear exponential decay with a fitted time constant of $\tau = 641(4)$ s, consistent with the previously reported lifetime[26] of $^{229m}$Th in $CaF_2$.

The long lifetime of the clock transition introduces a potential line shape asymmetry[26] when we measure with a 400 s laser on and 200 s laser off experiment cycle. Here, the excited state population from one data point does not fully decay when we start the next measurement. To avoid this asymmetry, we first measure a fine line shape by using the same irradiation cycle, but wait 1800 s, which is approximately three lifetimes, between adjacent frequency steps. This gives sufficient time for the $^{229m}$Th to decay to the ground state. For each step here, we keep the excitation frequency fixed during irradiation. Figure 3b shows one such measurement using this technique. A clear, symmetric line shape is observed and fit to a gaussian.

To use the scanning time more efficiently, we perform bidirectional scans for subsequent measurements. Here, we use the same 400 s on and 200 s off experiment cycle but scan the line shape once by increasing $f_{\mathrm{rep}}$ and another time by decreasing $f_{\mathrm{rep}}$. The scan range of each step is set to 100 kHz in the VUV. The two scans are each fit to gaussian line shapes, and the line center is determined by averaging the fit parameters from the two scans. Both line shapes have slight asymmetries due to the short detection window, but the asymmetries cancel each other when the two scan directions are averaged. We obtain line center uncertainties of ~4 kHz and full width at half maximum (FWHM) values of about 300 kHz. Fig. 3c shows a forward, backward, and averaged scan fit using this method. The plotted frequency point corresponds to the center frequency value of the scan range for each step. We use this bidirectional scan scheme for absolute frequency determination of the four peaks and for measuring the center frequency of a fifth peak from electric quadrupole splitting (see below).

**Comb mode determination**

While we illuminate the crystal with all the comb modes simultaneously, each spectroscopic feature corresponds to a nuclear excitation by one single comb mode. We use the technique previously demonstrated in Ref.[37] to determine the comb mode number exciting the transition and convert the $f_{\mathrm{rep}}$ (radio frequency) to the absolute frequency (VUV) of a given peak. We scan the same transition line with three different comb mode numbers $N_i$ (where $i$ denotes the scan number) by shifting $f_{\mathrm{rep}}$, as shown in Fig. 4a. These three distinct comb modes correspond to 100 kHz level jumps in $f_{\mathrm{rep}}$. The jump step sizes are chosen such that they are much greater than the uncertainty in the fitted line centers in absolute frequency, which is ~4 kHz. In combination with the bounds on the transition frequency set by Refs.[26,27], this allows us to determine the integer comb mode number exciting the line at each $f_{\mathrm{rep}}$ and thus the absolute frequency of the transition unambiguously. At each $f_{\mathrm{rep}}$ jump, we estimate the new $f_{\mathrm{rep}}$ position of each peak based on previous knowledge of the transition frequency and scan a small spectral region to find the exact position of the peak.

The uncertainty range of the nuclear transition at 2020408(3) GHz (given by the weighted average from Refs.[26,27]) corresponds to ~80 possible comb mode numbers $N_1 \approx 26848820$ to $26848900$ for $f_{\mathrm{rep}_1}$, shown on the $x$-axis of Fig. 4b. These mode numbers are used to compute initial guesses for the absolute frequency $\nu_{\mathrm{Th}}$. Assuming this frequency $\nu_{\mathrm{Th}}$ is correct, we use the fitted line centers at $f_{\mathrm{rep}_2}$ and $f_{\mathrm{rep}_3}$ to assign the closest integer comb mode $N_2$ and $N_3$. The comb equation can be written as:

$$\nu_{\mathrm{Th}_\mathrm{i}} = N_\mathrm{i} f_{\mathrm{rep}_\mathrm{i}} + 7 f_{\mathrm{CEO}}.$$

For the correct comb mode number assignment, the three $\nu_{\mathrm{Th}}$ from $f_{\mathrm{rep}_1}$, $f_{\mathrm{rep}_2}$, and $f_{\mathrm{rep}_3}$ should agree closely within uncertainty propagated from the fitted line center. We quantify this as the average frequency discrepancy:

$$\frac{1}{3}\sum_{i=1}^{3} |\nu_{\mathrm{Th}_\mathrm{i}} - \nu_{\mathrm{Th}_\mathrm{avg}}|$$

where $\nu_{\mathrm{Th}_\mathrm{i}}$ is the transition frequency calculated from $f_{\mathrm{rep}_\mathrm{i}}$ and $N_\mathrm{i}$, and $\nu_{\mathrm{Th}_\mathrm{avg}}$ is the weighted average of the three $\nu_{\mathrm{Th}}$. As seen in Fig. 4b, the optimal comb mode assignment produces the lowest frequency discrepancy of 10 kHz, which is within the 3-σ region calculated from the uncertainty of the fitted line center. A comb mode number assignment off by ±1 increases the average frequency discrepancy by an order of magnitude.

This optimal comb mode number assignment can also be visualized using a linear fit based on a rearranged comb equation:

$$\Delta N = (\nu_{\mathrm{Th}} - 7f_{\mathrm{CEO}})\Delta\frac{1}{f_{\mathrm{rep}}}$$

where $\Delta N$ and $\Delta\frac{1}{f_{\mathrm{rep}}}$ describe change in comb mode number and inverse fitted peak center from the highest repetition rate used, $f_{\mathrm{rep}_3}$. The two points corresponding to the two lower values of $f_{\mathrm{rep}}$, with error bars calculated from the uncertainties in the fitted line centers, are plotted in Fig. 4c. The insets provide an expanded view. The solid dark blue line passing through the origin with slope $\nu_{\mathrm{Th}_{\mathrm{avg}}} - 7f_{\mathrm{CEO}}$ corresponds to the optimal comb mode number assigned in Fig. 4b. The dashed green lines correspond to comb mode number assignments off by ±1, in clear disagreement with the measured data. The two analyses in Fig. 4b and 4c corroborate each other and demonstrate that we have assigned $N_{\mathrm{i}}$ with complete confidence.

With the comb mode numbers $N_{\mathrm{i}}$ for the three $f_{\mathrm{rep}}$ jumps uniquely determined for a given peak, we use the comb equation to convert from $f_{\mathrm{rep}}$ to absolute frequency. Fig. 4d shows the three scans of the same peak (individually normalized) and their gaussian fits plotted against absolute frequency, where the legend displays the determined comb number. We see the three scans overlap, confirming the correct determination of the absolute frequency.

This procedure is repeated for the four chosen peaks in Fig. 2 (highlighted in green). The absolute frequency is then calculated as the weighted average between the three scans for each peak, which are tabulated in Fig. 5. The left panel shows the four peaks (labelled a-d) in absolute frequency, individually normalized with the data and fits from $f_{\mathrm{rep}_3}$ shown. The center frequency of each line is determined to 4 kHz. The observed splittings are on the order of hundreds of MHz, which is shown in the top panel.

**Nuclear quadrupole structure**

The four chosen lines are attributed to the nuclear electric quadrupole structure. As the $^{229}$Th nuclei are embedded in a $CaF_2$ host, they experience a strong electric field gradient produced by the surrounding lattice ions and $F^-$ interstitials[47]. The interaction between the electric quadrupole moment $Q$ of the thorium nucleus and this electric field gradient gives rise to an electric quadrupole splitting, predicted to be on the order of hundreds of MHz[29,30]. The splittings can be extracted from diagonalizing the Hamiltonian[47,48]:

$$H_{\mathrm{E2}} = \frac{QV_{\mathrm{zz}}}{4I(2I-1)}\left[3{I_z}^2 - \boldsymbol{I}^2 + \eta\left(I_x^2 - I_y^2\right)\right]$$

where $Q$ is the spectroscopic nuclear electric quadrupole moment[17,25] in the ground ($Q_{\mathrm{g}}$) or isomeric state ($Q_{\mathrm{is}}$), and $\boldsymbol{I}$ is the nuclear angular momentum. The principal axis is chosen such that the electric field gradient matrix $V_{ij}$ is diagonal and is thus described by its $z$-component, $V_{\mathrm{zz}}$, and an asymmetry parameter, $\eta = (V_{\mathrm{xx}} - V_{\mathrm{yy}})/V_{\mathrm{zz}}$. The ground state of $^{229}$Th has a nuclear spin $I_{\mathrm{g}}$ = 5/2, while the isomeric excited state has $I_{\mathrm{is}}$ = 3/2. Transitions with $\Delta m = 0, \pm1$ are allowed by selection rules ($m = I_z$ is the $z$-axis angular momentum projection), though the asymmetry parameter $\eta$ leads to state mixing and therefore additional weak transitions. However, they are not observed in this work and do not correspond to the two weak peaks highlighted in blue in Fig. 2. The upper right panel of Fig. 5 shows the level diagram with the four absolutely determined frequencies assigned to their corresponding sublevel transitions a-d.

Using the above equation, we fit the measured line centers to the predicted quadrupole structure using $\eta$, $Q_g V_{zz}$, and $Q_{is} V_{zz}$ as the fit parameters. This procedure yields $\eta = 0.59163(5)$, $Q_g V_{zz} = 339.258(7)$ eb V/Å$^2$, and $Q_{is} V_{zz} = 193.387(5)$ eb V/Å$^2$ (1 eb $= 1.6022 \times 10^{-47}$C m$^2$ denotes one electron-barn). The extracted uncertainties are purely statistical. Independent of the crystal environment, the ratio of the quadrupole moments is extracted from the fitting procedure as $Q_{is}/Q_g = 0.57003(1)$, consistent with previously reported ratios[17,25] of 0.555(19) and 0.569(7). Taking previously published[49] value of $Q_g = 3.11(2)$ eb, this yields an electric field gradient $V_{zz} = 109.1(7)$ V/Å$^2$ for the $CaF_2$ crystal used in this experiment. In an upcoming publication, we will disentangle nuclear physics and material science by further exploring the quadrupole splittings as a function of material parameters.

In the initial full range scan shown in Fig. 2, we did not observe the fifth line of the quadrupole structure, corresponding to the $m_g = \pm 1/2$ to $m_{is} = \pm 3/2$ transition (labeled 'e' on the diagram). From the Clebsch-Gordan coefficients, this line is predicted to have only 1/10 the strength of the strongest line. Based on three of the known frequencies, we determine the expected absolute frequency of this fifth line using the sum rule, $\nu_e = \nu_c + \nu_d - \nu_a$ = 2 020 407 693.966(7) MHz. We repeat the scan over the corresponding $f_{rep}$ with our comb and indeed observe a weak line at 2 020 407 693.98(2) MHz. We determined this frequency with only a single $f_{rep}$ step using the expected comb mode number calculated from the expected line center.

Thus, the five lines corresponding to the nuclear quadrupole structure of the $^{229m}$Th transition have been measured and assigned, as shown in the Table of Fig. 5. With proper averaging of the quadrupole splitting patterns (ignoring higher order moments), we can recover the unsplit transition frequency free of the electrical field gradient. The transition frequency between the $I$ = 5/2 ground state and the $I$ = 3/2 excited state is determined as:

$$\nu_{Th} = \frac{1}{6}(\nu_a + 2\nu_b + 2\nu_c + \nu_d) = 2\,020\,407\,384\,335(2)\ \text{kHz}.$$

Hence, we report the frequency ratio between the $^{229}$Th nuclear clock transition and the $^{87}$Sr atomic clock to be:

$$\frac{\nu_{Th}}{\nu_{Sr}} = 4.707\,072\,615\,078(5)$$

for $^{229}$Th nuclei embedded in a $CaF_2$ host crystal at 150(1) K. Systematic uncertainty of this frequency ratio will be explored in future studies.

**Discussion**

Throughout the measurement campaign over two weeks, our measured line centers are consistent with each other within the measurement uncertainties, showing both the insensitivity of the $^{229}$Th nuclear transition frequency to external environment changes and the precise absolute frequency control of our laser system referenced to the JILA $^{87}$Sr clock. The experimentally measured FWHM of the nuclear resonance feature of around 300 kHz is most likely limited by the linewidth of our VUV frequency comb. During the cavity-enhanced harmonic generation process, the timing jitter of the fundamental laser is transferred to the upconverted light. Therefore, the $7^{th}$ harmonic has a $7^2 = 49$ times higher phase noise power spectral density compared to that of the fundamental[38]. This can lead to a substantial broadening of the comb linewidth. Using a low-noise reference laser to stabilize the fundamental comb could drastically reduce the VUV comb linewidth to the Hz level[38], leading to orders of magnitude improved uncertainty of the transition line center. This would allow us to probe the inhomogeneous broadening[29,30] in the host crystal, which is estimated to be a few hundred Hz.

Currently, the laser intensity (1 nW/comb mode focused to 0.2 mm diameter) and measured lifetime (641 s) correspond to a Rabi frequency of ~0.4 Hz, much smaller than the estimated inhomogeneous broadening[50]. Thus, to coherently control the nuclear state, a high-power VUV laser is needed. Now that the transition frequency has been precisely determined, a continuous wave laser with high output power can be built to address this transition. We note that efficient generation of VUV light in fiber systems[51] has also been demonstrated at kHz repetition rates, offering another possibility for scaling up the VUV comb power in a fiber-based system.

While an optical clock can in principle operate based on detecting fluorescence photons from the clock transition[30], novel approaches should be considered to accelerate the state detection. Akin to nuclear magnetic resonance, nuclear population in the ground and excited state manifold could be read out directly using radiofrequency nuclear quadrupole resonance spectroscopy[8], reducing the time needed for clock operation and thus achieving better clock stability. X-ray quenching effects[12,24] provide another means to accelerate the readout scheme.

Systematic shifts and broadenings of the nuclear transition frequency, such as those due to crystal temperature and local magnetic field, will be characterized in our future studies. With the systematic shifts under control, we will establish a time record for the frequency ratio of $\nu_{Th}/\nu_{Sr}$. Given the significantly enhanced sensitivity of the nuclear clock to fundamental constants[11], we expect to provide improved bounds on the temporal variation of the fine structure constant with moderately improved clock uncertainty.

To further improve the performance of a solid-state nuclear clock, important parameters of the crystal structure must be explored and understood. In particular, the origin of the two weak lines (Fig. 2, highlighted in light blue) in our full range scan remains unknown and could be coming from $^{229}$Th atoms doped in different charge compensation configurations within the crystal[47]. Studies comparing different host crystals[27], such as Th:$LiSrAlF_6$, could provide new insights. Differences in the electronic charge density and electric field gradient at the thorium nucleus in the $LiSrAlF_6$ crystal would lead to absolute frequency shifts and different splitting magnitude, respectively. These new measurements will provide important benchmarks for theoretical methods such as Density Functional Theory[47] and will lead to new nuclear physics insights.

High density quantum emitters with extremely long coherence times naturally provide a new system for quantum optics and quantum information studies. In our $^{229}$Th:$CaF_2$ target, there are $10^4$ atoms within a single $\lambda^3$ volume, where $\lambda$~148 nm is the transition wavelength. In pure $^{229}ThF_4$ crystals, the number density can reach $10^{22}$/cm$^3$, corresponding to $10^7$ atoms inside a single $\lambda^3$ volume, providing opportunities to study collective effects and nuclear quantum optics when inhomogeneous effects in solids are brought under control. Suppressing dephasing and harnessing quantum resources[52] in solid state $^{229}$Th targets will directly benefit the development of a solid-state portable clock and create tremendous opportunities for quantum/laser technology and material science.

In conclusion, we have established a direct frequency link between the $^{229m}$Th isomeric transition and the $^{87}$Sr atomic clock. We have improved the precision of the $^{229}$Th nuclear clock transition frequency by about six orders of magnitude and directly resolved the underlying nuclear quadrupole splitting. Our work provides a clear pathway forward to build and improve the nuclear clock. We herald the dawn of nuclear optical clocks that will enable numerous advances in fundamental physics, quantum physics, and precision measurement technologies.

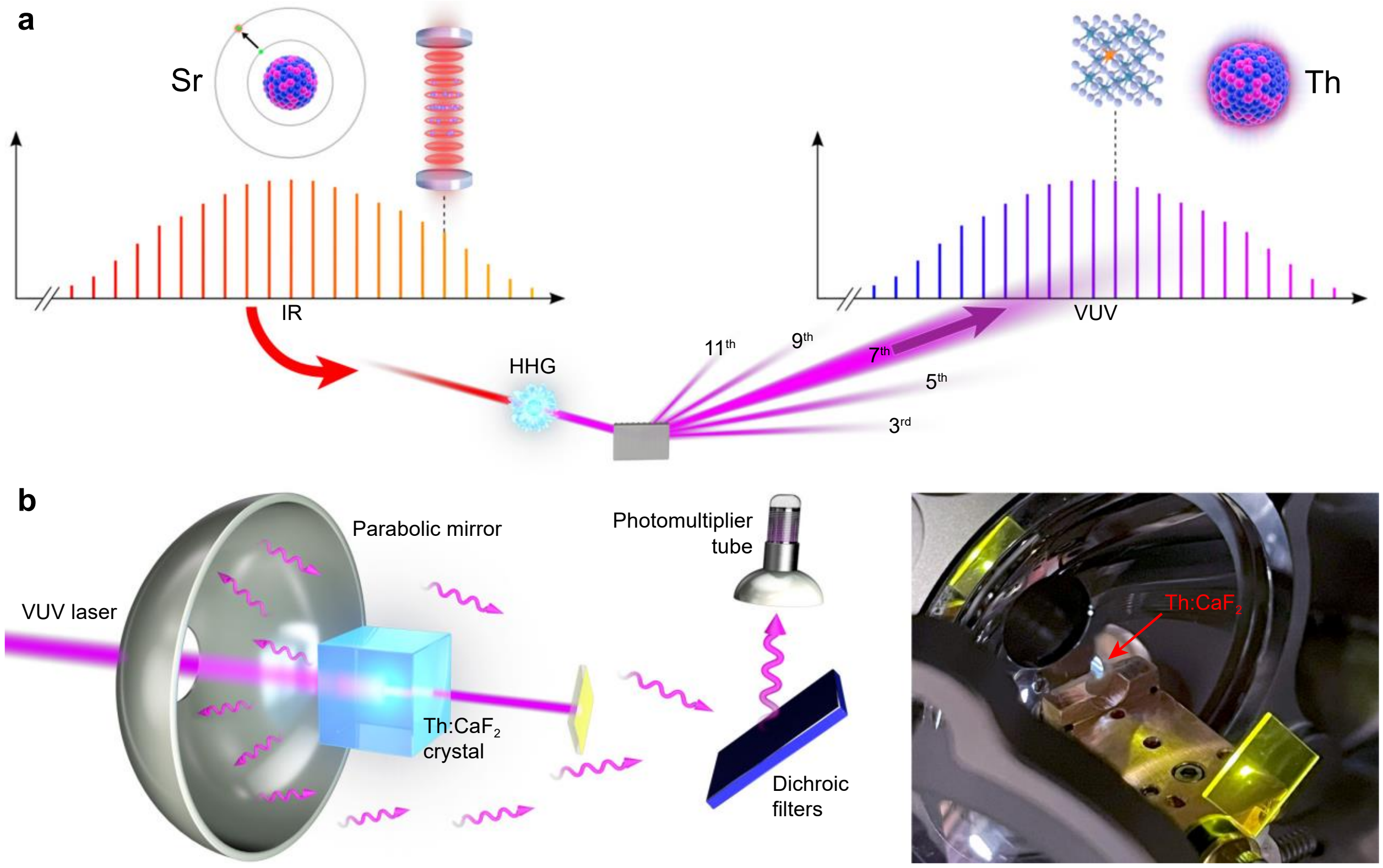

**Fig. 1 | VUV comb spectroscopy of the $^{229}$Th nuclear clock transition.** (a) An infrared frequency comb with a 75 MHz repetition frequency is stabilized by the JILA $^{87}$Sr optical lattice clock[2]. The infrared frequency comb is upconverted to VUV through a coherent, high-order harmonic generation (HHG) process inside a femtosecond enhancement cavity (not shown). A single VUV comb line in the 7$^{th}$ harmonic comb (~148.3 nm center wavelength) directly excites the $^{229}$Th nuclear clock transition, providing a direct frequency connection between the nuclear and atomic clocks. A representative crystal structure[53] of $^{229}$Th:$CaF_2$ is shown with the doped $^{229}$Th colored orange. (b) Illustration of the detection setup. After 400 s illumination, the excitation comb is shut off. Fluorescent photons from the decay of $^{229m}$Th embedded in the $CaF_2$ crystal are collected using a VUV-reflective parabolic mirror. A series of dichroic filters (only one shown) steer photons at the nuclear transition wavelength to a detector while suppressing background photons. A photomultiplier tube (PMT) is used to count individual VUV fluorescent photons for 200 s while the comb is off. (c) Photo of the Th:$CaF_2$ crystal under VUV irradiation. The Th:$CaF_2$ crystal scintillates, showing a visible white/blue trace of the laser path. A fluorescent screen (yellow) blocks the VUV comb transmission through the Th:$CaF_2$ crystal to protect the PMT from laser damage. The yellow fluorescence signal is also used for laser alignment and power monitoring.

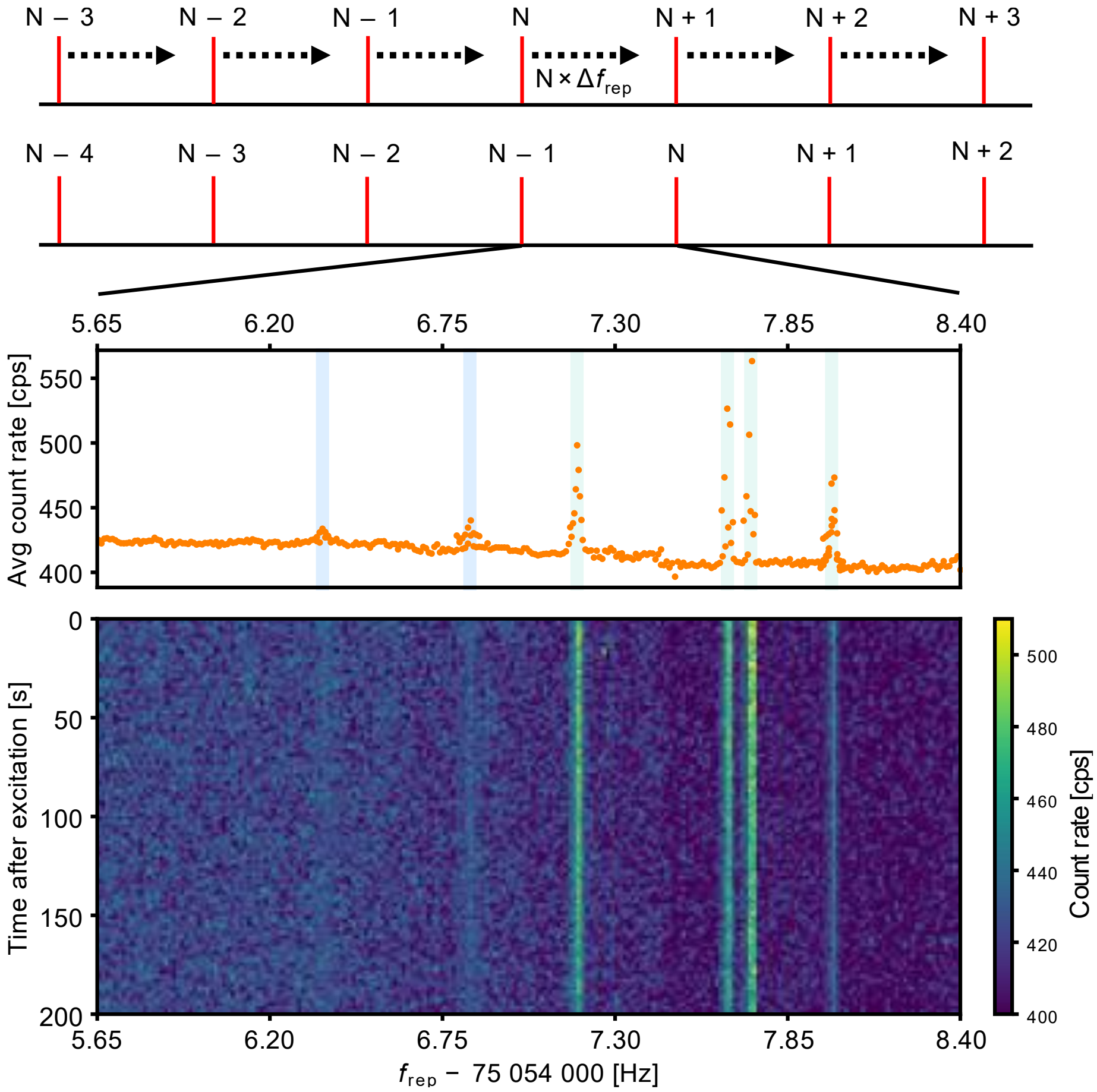

**Fig. 2 | A full range comb scan.** The VUV comb repetition rate is scanned by precisely tuning $f_{\text{rep}}$ to shift all the comb lines in parallel. The optical frequency gap between adjacent comb lines is fully covered when the $N^{\text{th}}$ comb mode overtakes the original frequency position of the ($N$+1) comb mode (top panel). Experimentally, the precise $f_{\text{rep}}$ control is achieved by fixing $f_{\text{CEO}}$, then digitally changing the phase lock offset frequency between the $^{87}$Sr clock laser and its nearest comb line. The average fluorescence photon count rate per second (cps) in the 200 s detection window is plotted against the comb repetition rate. Six distinct spectroscopic features are observed and highlighted in blue/green (middle panel). Four dominant peaks (green), which we later assign to electric quadrupole splittings of the nuclear transition, will be employed for absolute frequency determination. The count rate as a function of time (color bar, units in cps) at each frequency bin is plotted in the bottom panel, highlighting a nuclear excited state lifetime that is significantly longer than the detection window.

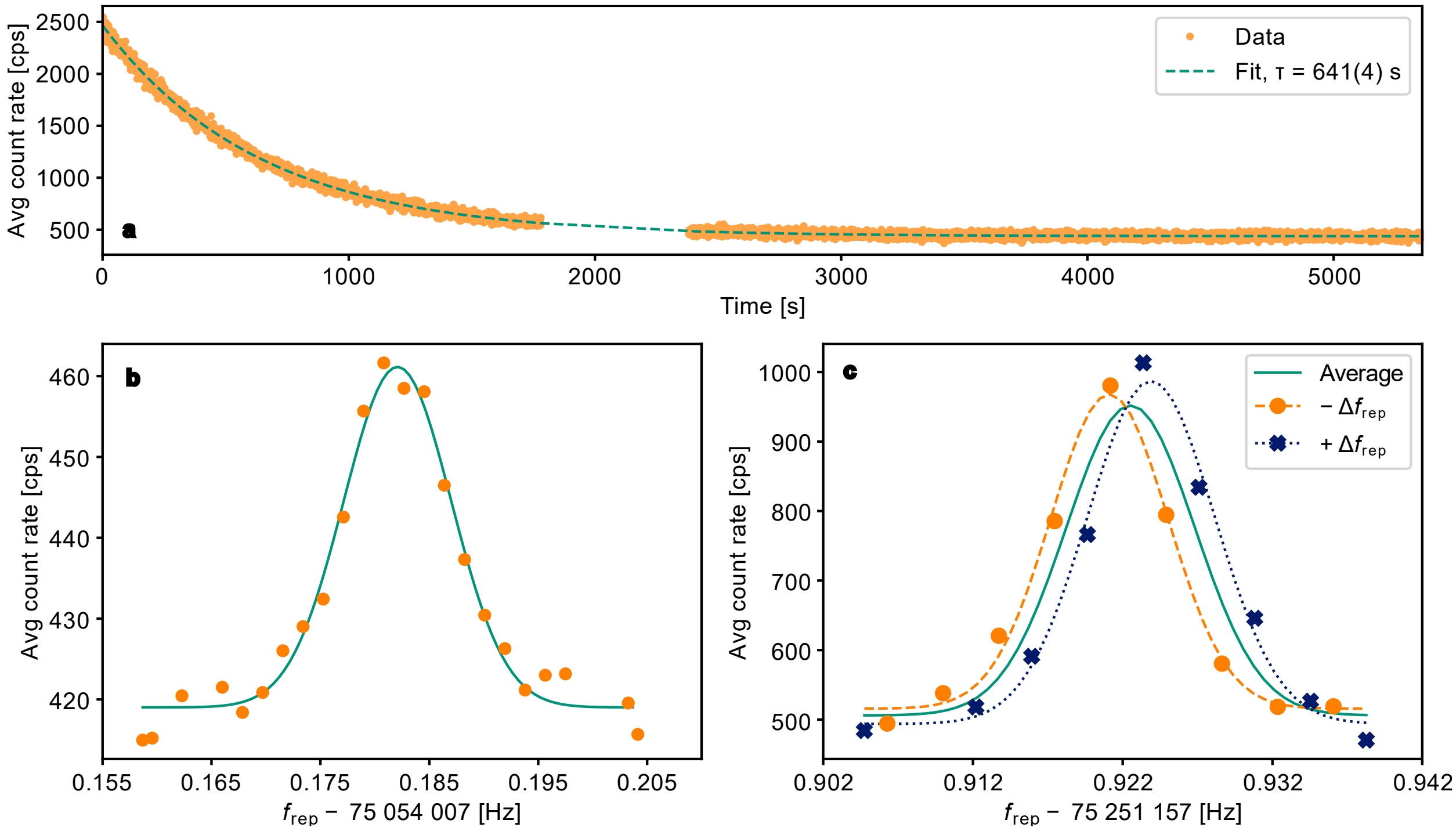

**Fig. 3 | Line shape and center frequency determination.** (a) Lifetime measurement of the excited clock state $^{229m}$Th ($m_g = \pm 5/2$ to $m_{is} = \pm 3/2$ transition, see Fig. 5, line b). The fluorescence photon count rate is monitored over time after laser excitation of 1200 s. A portion of the data around 2000 s is absent due to technical reasons. The experimental data is fit with a single exponential decay, leaving no structure in the fit residual. The extracted lifetime of the nuclear excited state is 641(4) seconds in the $CaF_2$ host crystal. (b) Fine scan showing the line shape (Fig. 5, line a) for one of the main clock transitions. We wait 1800 seconds in between each data point to avoid line shape asymmetry that could arise from residual excited population from previous scan steps. The fitted gaussian full width at half maximum is 0.0116(5) Hz in $f_{rep}$, corresponding to 310(10) kHz in absolute frequency. (c) Bidirectional scan (Fig. 5, line b). To accelerate the experimental cycle in line center determination for the four selected peaks, we use fast (400 s laser on and 200 s laser off) laser scan cycles. We perform scans by stepping $f_{rep}$ in both positive and negative directions, and their results, after proper intensity normalization, are averaged to eliminate systematic shifts caused by potential line shape asymmetries.

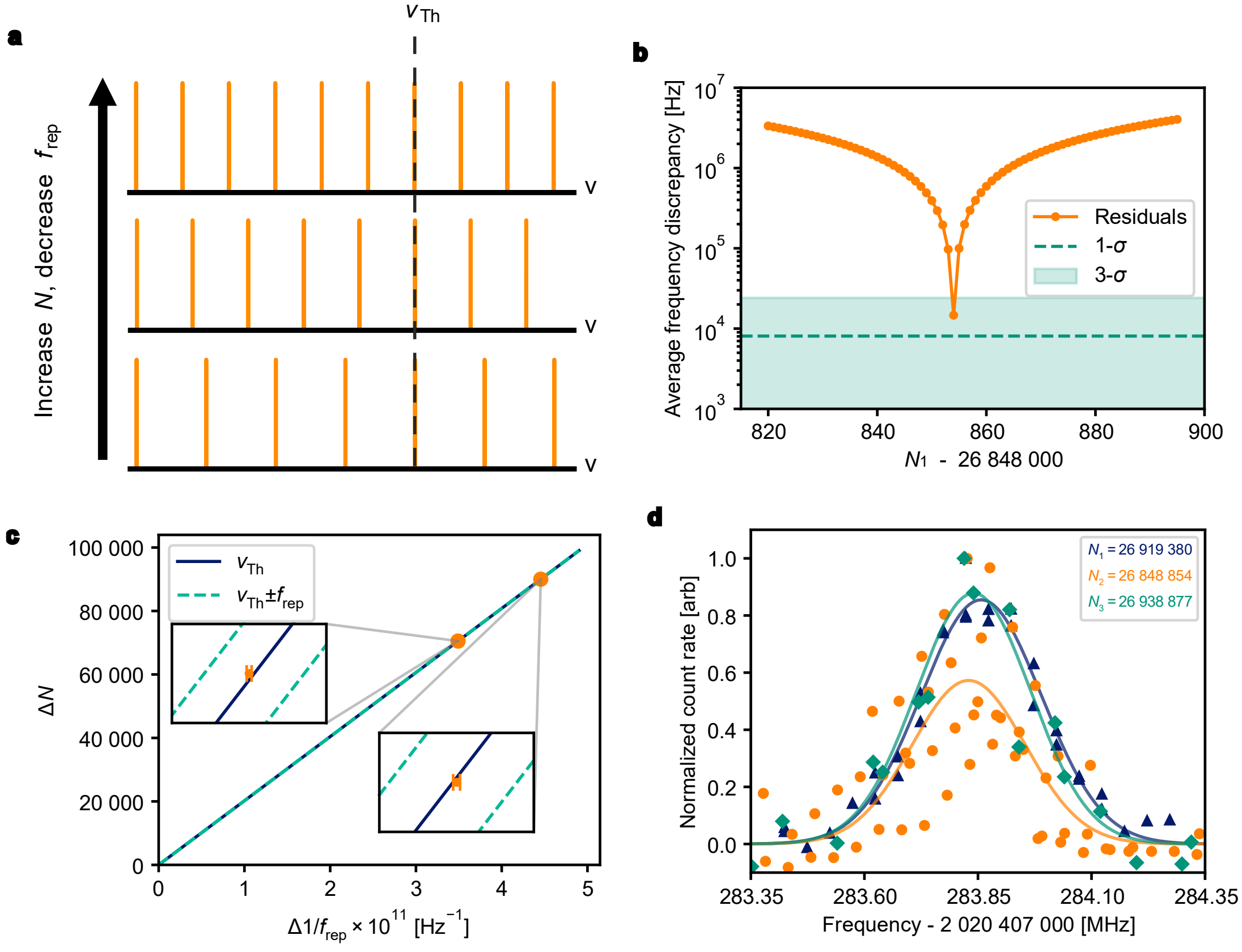

**Fig. 4 | Absolute frequency determination.** (a) Illustration for absolute frequency determination. We precisely scan the same transition line using three different values of $f_{rep}$. When these $f_{rep}$ values are sufficiently different from each other and are yet precisely known, the exact mode number associated with each $f_{rep}$ can be unambiguously determined[37]. (b) Previous measurements[26,27] of the nuclear transition energy provides an initial guess range of about 80 possible comb mode numbers. Integer trial values of comb mode number $N_1$ in the first scan are picked and the transition frequencies are calculated accordingly. We force the comb mode number for the subsequent two measurements to the closest integer according to $f_{rep}$ and calculate the transition frequency from them. The averaged frequency discrepancy between the measurements quantifies the comb mode assignment error. The lowest value of the discrepancy corresponds to the experimentally determined comb mode number assignment. The disagreement jumps by a factor of 10 when the comb mode is misassigned by ± 1. The indicated 1-σ line and 3-σ region corresponds to the statistical uncertainty of the fitted gaussian line center. (c) Another method for comb mode determination is to perform a linear fit between the mode number and the inverse value of $f_{rep}$. With an expanded view in the inset, we plot our measurement data for the two $f_{rep}$ jumps, with the corresponding 1-σ error bars. The solid dark blue line corresponds to the comb mode number assignment $N$ from (b), in agreement with the measurement. Dashed green lines correspond to comb mode $N \pm 1$, in clear disagreement with our measurements. (d) With the determination of the mode numbers, three scanned line shapes of a specific nuclear transition, corresponding to three different values of $f_{rep}$ are plotted together against their absolute optical frequencies, confirming their consistency.

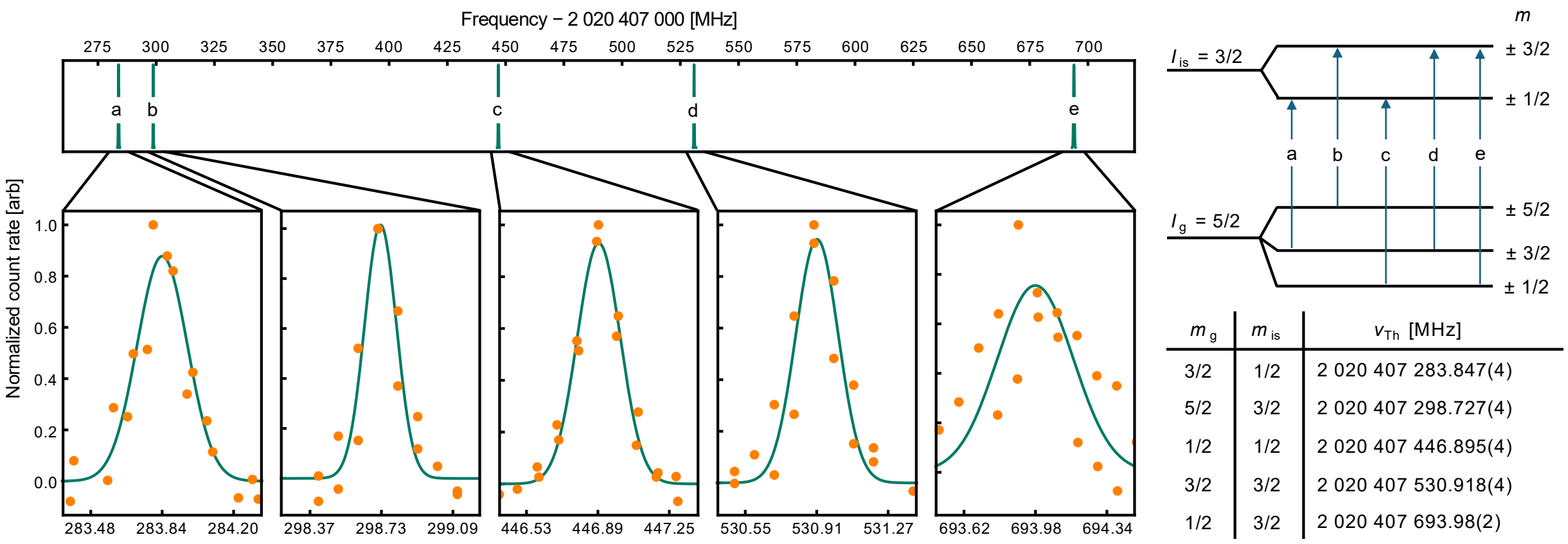

| $m_g$ | $m_{is}$ | $\nu_{Th}$ [MHz] |
|---|---|---|
| 3/2 | 1/2 | 2 020 407 283.847(4) |
| 5/2 | 3/2 | 2 020 407 298.727(4) |
| 1/2 | 1/2 | 2 020 407 446.895(4) |
| 3/2 | 3/2 | 2 020 407 530.918(4) |
| 1/2 | 3/2 | 2 020 407 693.98(2) |

**Fig. 5 | Direct spectroscopic measurement of nuclear electric quadrupole structure.** Top right panel shows the expected quadrupole line structure when $^{229}$Th is subject to an electric field gradient inside the $CaF_2$ crystal. With the absolute frequencies of line a, b, c, d determined by the direct comb spectroscopy technique, their corresponding quantum numbers are uniquely assigned. A zoomed view shows the measurement data for each line, with their relative intensity normalized to unity. The frequency of line e, which is a factor of 10 weaker in line strength, was calculated first using the relation $\nu_e = \nu_c + \nu_d - \nu_a$, and then confirmed with a comb scan using a single comb mode. Absolute frequencies of the five transition lines are listed in the table.

## Methods

### Frequency locking and scan parameters

We fully stabilize the frequency comb by locking it at two spectral points. First, we lock a specific comb mode to the $^{87}$Sr atomic clock. Second, we lock the carrier envelope offset frequency $f_{CEO}$ to a radiofrequency reference. The comb scan is accomplished by changing the offset frequency between one comb mode and the stable Sr clock laser (Extended Data Fig. 1). Briefly, supercontinuum light is generated in the pre-amplified comb via a highly nonlinear photonic crystal fiber (HNL PCF). Part of the supercontinuum light is also doubled using a periodically poled lithium niobate (PPLN) crystal. These two beams are beat against one another for the $f_{CEO}$ detection using the $f$–$2f$ scheme in an all-fiber setup[54]. We stabilize the $f_{CEO}$ by feeding back to the oscillator pump current. The lock gain sign and reference frequency are chosen such that we effectively stabilize the frequency of a virtual comb line corresponding to the comb $f_{CEO}$ to –8 MHz.

The supercontinuum comb light is also beat against the Sr clock laser at 698 nm to generate $f_{beat}$. We use the error signal from $f_{beat}$ to slowly steer the frequency of a narrow linewidth Mephisto laser centered at 1064 nm. The Mephisto laser is beat against the fundamental comb light. We stabilize the comb-Mehpisto beatnote by applying fast feedback to the comb oscillator cavity length using a piezo actuator along with an intra-oscillator electro-optical modulator. Scans are achieved by sweeping the $f_{beat}$ lock point with a DDS. Directly locking our comb to a narrow linewidth Mephisto laser near the fundamental spectrum reduces the comb phase noise and narrows the comb linewidth. The Mephisto laser is additionally coupled to the femtosecond enhancement cavity and is used to stabilize the cavity length using the Pound-Drever-Hall locking scheme[55]. An AOM is used to add an offset frequency to the Mephisto, allowing us to change the detuning between the frequency comb and cavity resonance. This comb-cavity detuning is used to mitigate cavity bistability caused by intracavity plasma nonlinearities[56,57].

### Crystal growth and properties

A thorium-doped calcium fluoride (Th:$CaF_2$) crystal is grown at TU Wien and cut (named "tiny-X2") from the ingot with the highest $^{229}$Th doping concentration ($5 \times 10^{18}$ /cm$^3$) to date. This ingot, "X2" (3.2 mm diameter, 10 mm long cylinder), is used in both the first laser excitation of the $^{229}$Th nucleus at PTB[26] and the $^{229}$Th population control at Spring-8[24]. Due to the radioactivity of $^{229}$Th, the crystal luminesces in the VUV and UV through Cherenkov radiation and annihilation of self-trapped-excitons, respectively[58]. Therefore, it is beneficial to match the crystal size to the excitation source spot size (0.2 mm) to maximize signal and reduce VUV and UV background. A corner of a X2 cylinder is cut (for methods, see Ref.[58]) using a 0.08 mm diamond wire saw such that the surface facing the laser has two edges at ninety degrees: 0.7 mm in width and 1.8 mm in height. The third edge of this surface is an arc with a radius of 1.6 mm. The depth of the crystal is 1.4 mm.

The original ingot "X2" is grown under vacuum using a miniaturized version of the vertical gradient freeze (VGF) method with single crystalline $CaF_2$ as seed. As described in detail in Ref.[44], the $^{229}$Th doping material is inserted in a pocket in the seed crystal before growth by preparing a $^{229}ThF_4$:$PbF_2$:$CaF_2$ precipitate as a carrier to facilitate physical handling of the <1 mg of $^{229}$Th. In the VGF procedure, the top part of the seed melts, and subsequently a single crystalline Th:$CaF_2$

is grown as the melt is slowly cooled. The $PbF_2$ acts as oxygen scavenger and evaporates. During growth, the radioactivity of $^{229}$Th induces radiolysis in the melt which causes evaporation of $F_2$, resulting in a fluoride-deficient non-stoichiometric Th:$CaF_2$ crystal. The fluoride deficiency causes formation of Ca metallic inclusions, which through Mie-scattering[59] cause a reduction of the crystal transparency at the isomer wavelength. Using superionic fluoride transfer, as described in Ref.[45], fluoride is added to the crystal efficiently without losing the single crystal structure, recovering VUV transparency. The crystal properties of tiny-X2 are shown in Extended Data Table 1.

**Spectroscopy system alignment**

The thorium-doped $CaF_2$ crystal is glued onto a $MgF_2$ plate and mounted inside the vacuum chamber. A liquid nitrogen dewar is installed to the top of the chamber, which is connected to the crystal mount via a cold finger and copper thermal link. The link is attached in a configuration that would minimize fluorescent photon loss. A temperature sensor attached to the mount monitors the crystal temperature during scan operation. The operating temperature remains stable at ~150(1) K during all scans.

Fig. 1B (right) shows proper alignment of the VUV comb to the crystal. The blue fluorescence from the crystal serves as a monitor of the VUV power and an alignment check. When the beam is aligned correctly, a blue fluorescent line can be seen passing through the entire crystal. The VUV beam is terminated at a Ce:YAG scintillation plate for alignment and power monitoring, and to avoid damage of the PMT downstream in the optical beam path.

The crystal is placed at the focal point of the VUV reflective parabolic mirror. The mirror has a 12 mm diameter hole in the vertex to allow the VUV beam to pass through. Fluorescent photons from the crystal are collimated by the parabolic mirror with a geometrical collection efficiency[60] of approximately 0.4. The photons are spectrally filtered by three dichroic mirrors and focused via a $MgF_2$ lens to the PMT.

**Data availability**

The data that support the findings of this study are available from the corresponding author upon appropriate request.

**Additional references (Methods)**

**Acknowledgements**
We would like to thank Kyungtae Kim, Alexander Aeppli, William Warfield, and William Milner for building and maintaining the JILA $^{87}$Sr optical clock, Dahyeon Lee, Zoey Hu, and Ben Lewis for building and maintaining the JILA stable laser and the cryogenic Si cavity, the entire crystal growth team at TU Wien for preparation of the thorium-doped crystal, Martin E. Fermann and Jie Jiang for help in constructing the high power IR frequency comb, Kim Hagen, Calvin Schwadron, Kyle Thatcher, Hans Green, Danny Warren, and James Uhrich for help in designing and building mechanical parts used in the detection setup, Terry Brown and Ivan Rýger for help in designing and making electronics used in the experiment, Margaret Ashton, Bradly C. Denton, and McKenzie R. Statham for help in shipment of radioactive samples, and Eric Hudson, Ekkehard Peik, Joonseok Hur, James Thompson, Johannes Weitenberg, and Akira Ozawa for helpful discussions. We also thank IMRA America, Inc. for collaboration.

We acknowledge funding support from Army Research Office (W911NF2010182); Air Force Office of Scientific Research (FA9550-19-1-0148); National Science Foundation QLCI OMA-2016244, National Science Foundation PHY-2317149; and National Institute of Standards and Technology. J.S.H. acknowledges support from the National Research Council Postdoctoral Fellowship. L.v.d.W. acknowledges funding from the Humboldt Foundation via a Feodor Lynen fellowship. P.G.T. acknowledges support from the European Research Council (ERC) (Horizon 2020, No. 856415) and the European Union's Horizon 2020 program (Grant No. 664732).

The $^{229}$Th:$CaF_2$ crystal was grown in TU Wien with support from European Research Council (ERC) (Horizon 2020, No. 856415) and the Austrian Science Fund (FWF) [Grant DOI: 10.55776/F1004, 10.55776/J4834, 10.55776/ PIN9526523]. The project 23FUN03 HIOC [Grant DOI: 10.13039/100019599] has received funding from the European Partnership on Metrology, co-financed from the European Union's Horizon Europe Research and Innovation Program and by the Participating States. We thank the National Isotope Development Center of DoE and Oak Ridge National Laboratory for providing the Th-229 used in this work.

L.v.d.W.'s present address is Johannes Gutenberg-Universität Mainz, Institut für Physik, Staudingerweg 7, 55128 Mainz, Germany. K.B.'s present address is Laboratory for Ultrafast Microscopy and Electron Scattering (LUMES), Institute of Physics, École Polytechnique Fédérale de Lausanne (EPFL), Lausanne CH-1015, Switzerland.

**Author contributions**
C.Z., T.O., J.S.H., J.D., L.v.d.W., K.B., T.S., and J.Y. conceived and planned the experiment, K.B., A.L., G.A.K., and T.S. grew the thorium-doped crystal and characterized its performance, P.G.T. provided valuable insight and the parabolic mirror, C.Z., T.O., J.S.H., J.D., L.v.d.W., P.L., and J.Y. performed the measurement and analyzed the data. All authors wrote the manuscript.

**Competing interests**
The authors declare no competing interests.

**Additional information**
Correspondence and requests for materials should be addressed to Chuankun Zhang and Jun Ye.

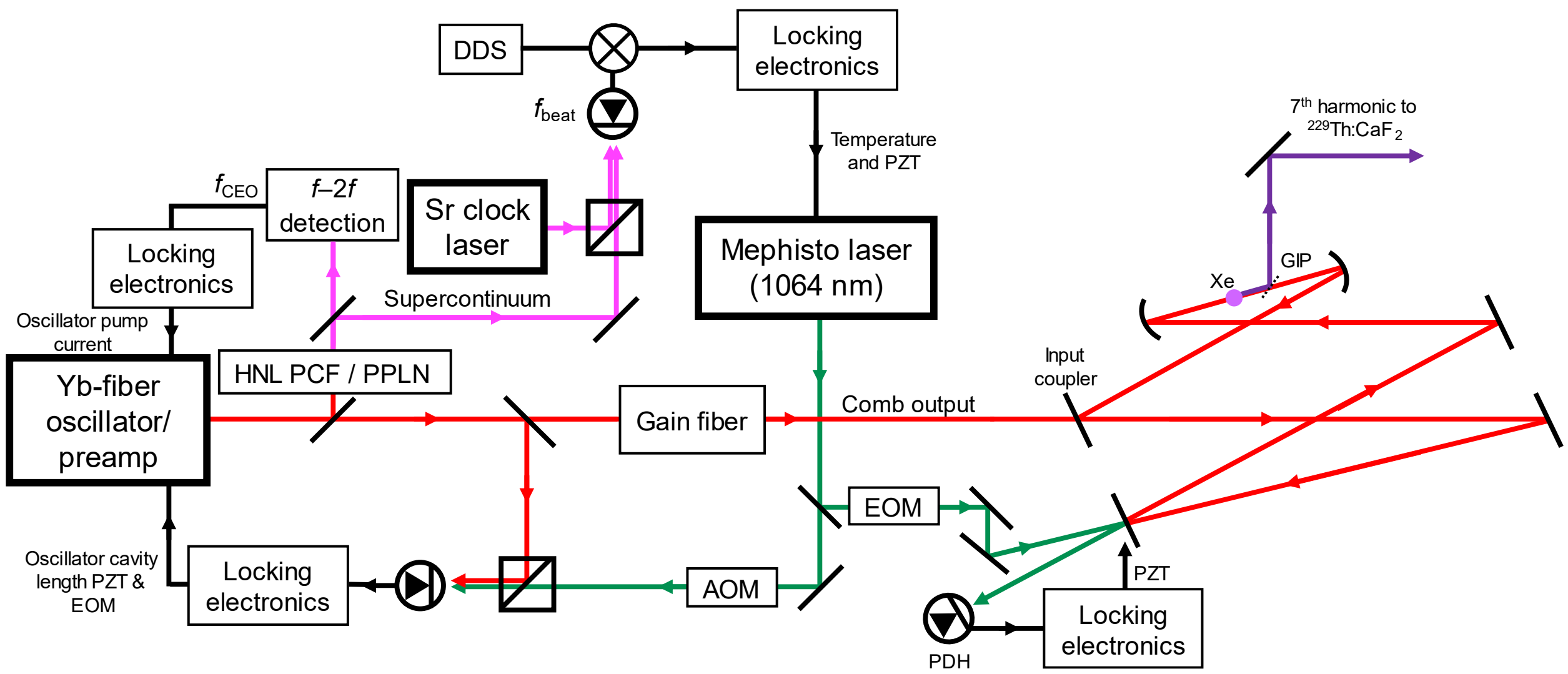

**Extended Data Fig. 1 | Locking scheme used in our experimental setup.** A Yb-fiber oscillator is used to generate the fundamental frequency comb[40]. The light is amplified using a chirped pulse amplification scheme in a large mode area gain fiber. The output comb light (average power 40-50 W) is coupled to a femtosecond enhancement cavity with finesse ~600 to further enhance the peak power for efficient cavity-enhanced high harmonic generation. The 7th harmonic is outcoupled using a grazing incidence plate[42,43] (GIP) and directed to the sample chamber. A portion of the pre-amplified comb light is picked off and focused to a highly nonlinear photonic crystal fiber (HNL PCF) for broadband supercontinuum generation. The light is also doubled using a periodically poled lithium niobate (PPLN) crystal. These two beams generate a beatnote that directly reports on $f_{CEO}$ ($f$–$2f$ detection), which can be fed back to the pump current for $f_{CEO}$ locking. The supercontinuum light is beatnote locked against the Sr clock light at 698 nm through an auxiliary narrow linewidth Mephisto laser at 1064 nm. The beatnote $f_{beat}$ is mixed with a DDS output and is used to steer the Mephisto laser frequency. The Mephisto output is passed through a fiber acousto-optic modulator (AOM) to generate a frequency offset and is beat against a portion of the preamplified fundamental comb light. The control signal is fed back to the oscillator cavity length to close the loop for the $f_{beat}$ lock. We conduct our scans by changing the DDS offset frequency, which ultimately changes the comb repetition frequency without shifting $f_{CEO}$. An additional portion of the Mephisto light is picked off and modulated with an electro-optical modulator (EOM) for Pound-Drever-Hall locking of the enhancement cavity. The offset between the locked cavity resonance and the fundamental frequency comb can be tuned by adjusting the AOM offset frequency to mitigate intracavity plasma instabilities[56,57]. PZT, piezo-electric actuator.

| Base area | 0.823 $mm^2$ |
|---|---|
| Weight | 3.7 mg |
| Column density | $8 \times 10^{15}$ $mm^{-1}$ |
| Activity | 18.1 kBq |
| Transmission | 45% at 150 nm |

**Extended Data Table 1 | Properties of Tiny-X2 crystal.** Physical dimensions and properties of the Tiny-X2 crystal are listed in the table.